\documentclass[a4paper,UKenglish]{lipics-v2021}

\hideLIPIcs  

\title{String Diagrams for Monoidal Categories, in Rocq}

\author{Damien Pous}{Plume team, CNRS, LIP, ENS de Lyon, UMR 5668, France \and \url{http://perso.ens-lyon.fr/damien.pous/} }{Damien.Pous@ens-lyon.fr}{https://orcid.org/0000-0002-1220-4399}{}

\authorrunning{D. Pous}
\Copyright{Damien Pous}
\ccsdesc[500]{Theory of computation~Logic}

\keywords{Monoidal categories, string diagrams, proof assistants, graphical proofs, Rocq}

\supplement{\url{https://perso.ens-lyon.fr/damien.pous/string-diagrams/}}

\nolinenumbers 

\EventEditors{John Q. Open and Joan R. Access}
\EventNoEds{2}
\EventLongTitle{42nd Conference on Very Important Topics (CVIT 2016)}
\EventShortTitle{CVIT 2016}
\EventAcronym{CVIT}
\EventYear{2016}
\EventDate{December 24--27, 2016}
\EventLocation{Little Whinging, United Kingdom}
\EventLogo{}
\SeriesVolume{42}
\ArticleNo{23}

\usepackage{harengs}

\begin{document}

\maketitle

\begin{abstract}
  We present a Rocq library for monoidal categories, which includes a decision procedure for proving equality of morphisms as well as notations that make it possible to reason as if they were strict, inferring MacLane isomorphims automatically in the background.
  Together with an external tool for visualising and editing string diagrams, this make it possible to perform rewriting steps in monoidal categories graphically, and to translate them into textual formal proofs which are concise and readable.
\end{abstract}

\section*{Introduction}
\label{sec:intro}

Modern proof assistants made it possible to formalise impressive results in both mathematics and computer science. Nevertheless, besides situations where the considered proofs involve numerous computations which would hardly be done by a human-being (e.g., Kepler's conjecture or the four-colour theorem), proof assistants often make it harder to write convincing proofs: proofs that we can read and get inspiration from. This is particularly true in domains where scientists have developed notational systems that go beyond traditional one-dimensional text to convey their ideas.

One such domain is category theory: proofs in category theory are often best displayed as commutative diagrams, where we get to see that the whole diagram commutes because all of its inner components do.
This notational system has its limitations: when the diagram of interest is not planar, one has to be ingenuous to draw it on paper in an intelligible way. Still, despite recent efforts to design prover interfaces to output proof states and input user instructions graphically~\cite{lafontDiagramEditorMechanise2024,chabassierGraphicalInterfaceCategory2025,chabassierAspectsCategoryTheory2025}, it remains difficult to formalise these proofs in such a way that the resulting object faithfully conveys the initial idea.

In this paper we focus on a specific concept of category theory: monoidal categories~\cite{Benabou63,MacLane63,Kelly64,EilenBergKelly66,Benabou72,MacLane:CWM}.
Those are categories with an associative bifunctor, thus a fairly general concept, which was subsequently refined into many variations (braided, symmetric, Cartesian, closed).
Still, monoidal categories alone already make it possible to study concrete concepts such as monads: ``a monad is just a monoid in the monoidal category of endofunctors''.
Moreover, monoidal categories come with a neat notation system: \emph{string diagrams}~\cite{Hotz65,Penrose71,Kelly72,Penrose_Rindler_1984,JOYAL91,Street96}. String diagrams are not commutative diagrams as mentioned above: they are pictures that make it possible to represent morphisms in monoidal categories, abstracting over irrelevant details induced by textual presentations of these morphisms. See~\cite{Piedeleu_Zanasi_2025} for a recent introduction.

We present a Rocq library for monoidal categories, together with a graphical editor for string diagrams.
Combined together, they make it possible to perform reasoning steps graphically, at the level of string diagrams, and to translate those graphical steps into clear and concise formal ones. 

This is made possible by developing three automation tactics: a first one to infer ``MacLane isomorphisms'' automatically, so that the user does not need to write them explicitly; a second one to exploit the unicity of these morphisms (which is derived from the coherence conditions of monoidal categories); and a third one to prove general morphism equalities.

We first briefly recall standard categorical notation (Section~\ref{sec:cat}) before discussing monoidal categories (Section~\ref{sec:mcat}) and string diagrams (Section~\ref{sec:sd}).
Then we present our diagram editor (Section~\ref{sec:gph}), illustrating its usage on a simple example: the composition of two monoids. We finally describe the Rocq library (Section~\ref{sec:rocq}), which makes it possible to follow graphical steps and produce convincing formal proofs. We conclude with related works (Section~\ref{sec:rw}) and directions for future work (Section~\ref{sec:fw}).

\section{Categorical notation}
\label{sec:cat}

We assume some familiarity with the basic concepts of category theory: morphisms, isomorphisms, functors, and natural transformations~\cite{MacLane:CWM,borceuxHandbookCategoricalAlgebra1994}.
We use letters $A,B\dots$ to range over the objects of categories. Given two such objects, we write $f\colon A\homto B$ when $f$ is a morphism from $A$ to $B$, and $f\colon A\simeq B$ when $f$ is an isomorphism from $A$ to $B$.
Given two morphisms $f\colon A\homto B$  and $f\colon B\homto C$, we write $f\semi g$ for their composition, which is a morphism from $A$ to $C$. We prefer this notation to the reversed one, $g\circ f$, as it makes the back and forth translations with diagrams more natural. In contexts where a morphism is expected, we use $A$ to denote the identity morphism on $A$.

\section{Monoidal categories}
\label{sec:mcat}

A \emph{monoidal category} is a category $\C$ together with
\begin{itemize}
\item a bifunctor $\otimes\colon \C\times\C\to \C$ called the \emph{tensor}, whose application to two morphisms $f,g$ is written $f\cdot g$;
\item an object $1\colon\C$ called the \emph{unit};
\item and three natural families of isomorphisms
  \begin{align*}
    \tag{associator}
    \alpha_{A,B,C}\colon &(A\otimes B)\otimes C \simeq A\otimes (B\otimes C)\\
    \tag{left unitor}
    \lambda_{A}\colon &1\otimes A \simeq A\\
    \tag{right unitor}
    \rho_{A}\colon &A\otimes 1 \simeq A
  \end{align*}
\item such that the following \emph{coherence} diagrams commute:
  \begin{gather}
    \label{ax:triangle}
    \tag{triangle}
    \begin{tikzcd}[ampersand replacement=\&]
      (A\otimes 1)\otimes B
      \arrow[rr,"{\alpha_{A,1,B}}" below]
      \arrow[rd,"{\rho_A\cdot B}" below left]
      \&\& A\otimes (1\otimes B)
      \arrow[ld,"{A\cdot\lambda_B}"]\\
      \& A\otimes B
    \end{tikzcd}\\
    \label{ax:pentagon}
    \tag{pentagon}
    \begin{tikzcd}[ampersand replacement=\&]
      ((A\otimes B)\otimes C)\otimes D
      \arrow[rrrr,"{\alpha_{A\otimes B,C,D}}" below]
      \arrow[d,"{\alpha_{A,B,C}\otimes D}" left]
      \&\&\&\&(A\otimes B)\otimes (C\otimes D)
      \arrow[d,"{\alpha_{A,B,C\otimes D}}" right]
      \\
      (A\otimes (B\otimes C))\otimes D
      \arrow[rr,"{\alpha_{A,B\otimes C,D}}" below]
      \&\&A\otimes ((B\otimes C)\otimes D)
      \arrow[rr,"{A\otimes\alpha_{B,C,D}}" below]
      \&\&A\otimes (B\otimes (C\otimes D))
    \end{tikzcd}    
  \end{gather}
\end{itemize}

Every category with finite products is monoidal, by taking the Cartesian product for the tensor and the terminal object for the unit. Endofunctors on a category form a monoidal category, where morphisms are natural transformations, tensor product is functor composition and unit is the identity functor.
The latter monoidal category is \emph{strict}, in the sense that the associators, left unitors, and right unitors consist of identity morphisms (in which case the two coherence diagrams hold trivially). The former is not strict in general: products are defined only up to isomorphisms.

\subsection{Laws}
\label{ssec:laws}

Besides the last two coherence diagrams, whose nature is rather bureaucratic, the above definition implicitly provides some important laws in monoidal categories.

First, we have bifunctoriality of tensor product: for all morphisms $f,g,f',g'$ of appropriate types, and all objects $A,B$, we have
\begin{align*}
  (f\semi g) \cdot (f'\semi g') &= f\ccdot f'\ssemi g\ccdot g' &
  A\cdot B &= A\otimes B\enspace.
\end{align*}
(Note that $(\cdot)$ binds stronger than $({;})$ in expressions, and that the second equation is about identity morphisms: $A\cdot B$ is the tensor product of the identity morphisms on $A$ and $B$, and $A\otimes B$ is the identity morphism on that object.)
As a consequence, for all morphisms $i\colon I\homto I'$ and $i\colon J\homto J'$ we have the following exchange law:
\begin{align*}
  I\ccdot j \ssemi i\ccdot J' = i\cdot j = i\ccdot J \ssemi I'\ccdot j\enspace.
\end{align*}
The above laws are used frequently, typically to bring together subexpressions which are far apart in some textual writing (e.g. $i$ and $j$ in the right-hand side of the last one), so that we can use some further knowledge about these subexpressions together.
We shall see in the sequel that while the one dimensional textual writing of morphisms makes this tedious, string diagrams, which are two dimensional, make it possible to detect such potential rewritings visually and much more easily.

Second, we have naturality of the associators and unitors: for all morphisms $f\colon A\homto A'$,
$g\colon B\homto B'$, and $h\colon C\homto C'$, we have:
\begin{align*}
  (f\cdot g)\cdot h \ssemi \alpha_{A',B',C'} &= \alpha_{A,B,C}\ssemi f\cdot (g\cdot h)\\
  1\cdot f \ssemi \lambda_{A'} &= \lambda_{A}\semi f\\
  f\cdot 1 \ssemi \rho_{A'} &= \rho_{A}\semi f\enspace.
\end{align*}
Given that $\alpha,\lambda$, and $\rho$ are isomorphisms, these laws can also be written as follows:
\begin{align*}
  (f\cdot g)\cdot h  &= \alpha_{A,B,C}\ssemi f\cdot (g\cdot h) \ssemi \alpha_{A',B',C'}^{-1}\\
  1\cdot f &= \lambda_{A}\semi f \semi \lambda_{A'}^{-1}\\
  f\cdot 1 &= \rho_{A}\semi f \semi \rho_{A'}^{-1}\enspace.
\end{align*}

\subsection{Strict monoidal categories?}
\label{ssec:strict}

Strict monoidal categories are much easier to work with, as we do not have to care about associators and unitors. For instance, the last three laws just become:
\begin{align*}
  (f\cdot g)\cdot h  &= f\cdot (g\cdot h)\\
  1\cdot f &= f\\
  f\cdot 1 &= f\enspace.
\end{align*}
On paper, when proving that a diagram holds, we may ``pretend'' that the considered monoidal category is strict, thanks to \emph{strictification}: every monoidal category is (monoidally) equivalent to a strict one~\cite{MacLane:CWM}. This technique is frequently used in the literature, when monoidal categories are not just assumed to be strict, ``for the sake of simplicity''. 


Unfortunately, these two approaches are not helpful in a type-theory based formalisation.
Indeed, strict monoidal categories are not easier than the general ones in such a setting.
Imagine for instance that we replace the associator by a family of propositional equalities:
\begin{align*}
  a\colon \forall A,B,C,~(A\otimes B)\otimes C = A\otimes (B\otimes C)\enspace.
\end{align*}
Given three morphisms $f,g,h$, we expect $(f\cdot g)\cdot h = f\cdot (g\cdot h)$ to hold in a strict monoidal category. However, this equation is not even well-typed: it relates morphisms whose sources and targets are not definitionally equal. They are propositionally equal, though, thanks to $a$, and we can use an explicit ``casting'' operation\footnote{We use the terminology from programming languages there: we cast a value of a given type into an other type; casts are also called \emph{transports} in proof relevant contexts.} to state this equation:
\begin{align*}
  (f\cdot g)\cdot h = \mathord{cast}\, (a \_\,\_\,\_)\, (a \_\,\_\,\_)\, (f\cdot (g\cdot h))\enspace.
\end{align*}
Such an equation actually corresponds to naturality of $\alpha$, and following this path, we reach very quickly the point where we need coherence conditions on cast operations that correspond to the coherence diagrams of non-strict monoidal categories. All in all, the induced overhead is equivalent to working with a general monoidal category in the first place.

Note however that some monoidal categories are inherently strict, even in type theory.
For instance, assuming extensionality, the category of endofunctors is strict monoidal, definitionally (e.g., given three functors $F,G,H$, the compositions $(F\circ G)\circ H$ and $F\circ(G\circ H)$ are definitionally equal.)
While this makes it easier to work in such concrete monoidal categories, our point is that formalising strict yet abstract monoidal categories in type theory would not be simpler than formalising general ones---even in a context with univalence, where isomorphisms would boil down to propositional equalities: we would still need to deal with explicit coherence conditions on transports~\cite{Ahrensetal2021}.

Although we formalise general monoidal categories in the present work, we provide support that make it possible to reason as if they were strict: this is one of our main contributions.

\subsection{Formalisation}
\label{ssec:rocqcat}

We started from the Rocq formalisation of categories from~\cite{ArsacHP26}. This library implements $\mathcal E$-categories~\cite{DubricDybjierScott98:pcategories,Kinoshita97:ecategories}, where each homset is equipped with a user-defined equivalence relation~$(\equiv)$, so that we do not need axioms such as functional extensionality or proof-irrelevance. It is structured using \emph{Hierarchy Builder (HB)}~\cite{cohenHierarchyBuilderAlgebraic2020}, which provides a high-level language to define hierarchies of structures, implementing \emph{packed classes}~\cite{garillotPackagingMathematicalStructures2009} in the background.

We slightly extended this library to encompass basic properties of natural isomorphisms and products of categories, so that we can easily obtain bifunctors. We also added some basic automation tactics, by rewriting and reflection, to develop subsequent proofs more easily.

The system of notations and coercions makes it possible to keep compact statements that follow closely the conventions we use in the present paper.



% ...


\section{String diagrams}
\label{sec:sd}

String diagrams are the right abstraction for morphism expressions in monoidal categories: they make it possible to abstract away the notational clutter induced by terms.
We will not need a formal definition of string diagrams, even in Rocq, so that we only provide an intuitive one here.

A \emph{string diagram} in a monoidal category is a graph, which consists of:
\begin{itemize}
\item a sequence of \emph{outer input ports}, labelled by objects;
\item a sequence of \emph{outer output ports}, labelled by objects;
\item a set of \emph{nodes}, labelled by morphisms and placed in the plane;
\item a set of \emph{edges}.
\end{itemize}
When a node is labelled by a morphism $f\colon A_1\otimes\dots\otimes A_n\homto B_1\otimes\dots\otimes B_m$ (whichever the concrete parenthesising), it defines $n$ \emph{inner output ports} and $m$ \emph{inner input ports}, labelled by $A_1,\dots, A_n$ and $B_1,\dots,B_m$, respectively. Edges are directed and they always connect an input port with a given label to an output port with the same label.
In addition, all ports should be incident to exactly one edge, the graph should be acyclic, and its placement in the plane be a planar embedding: edges cannot cross. String diagrams are considered up to continuous deformations of their embedding in the plane: nodes can be moved along, as long as they remain in the same face of the graph.

\begin{figure}
  \centering
  \begin{align*}
    \vcenter{\hbox{\includegraphics[scale=.5]{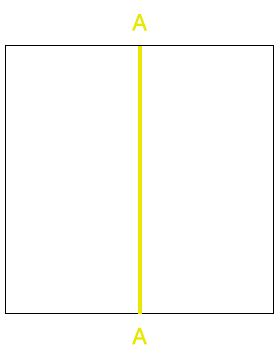}}}&\qquad\qquad&
    \vcenter{\hbox{\includegraphics[scale=.4]{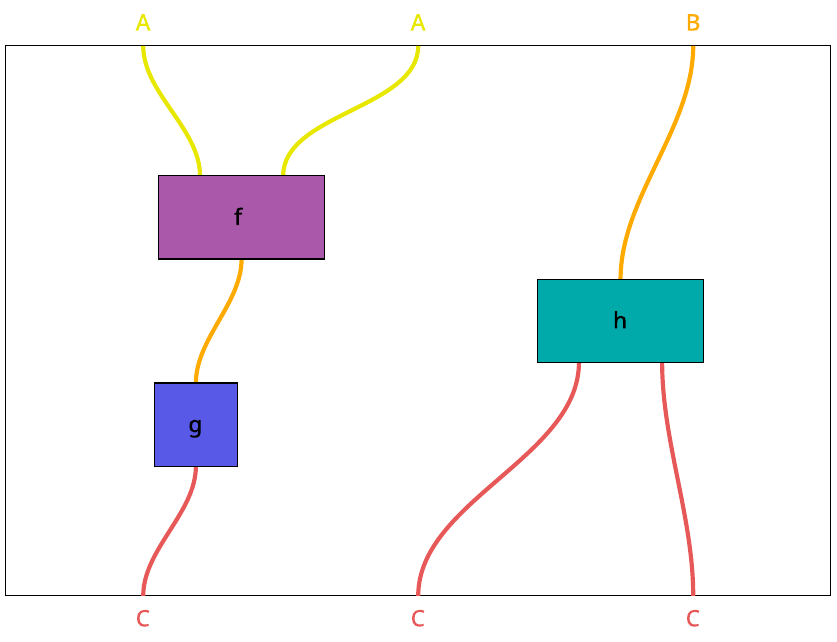}}}
  \end{align*}
  \caption{String diagrams corresponding to the identity on $A$ and to the expression $(f\semi g)\cdot h$ with $f\colon A\otimes A\homto B$, $g\colon B\homto C$ and $h\colon B\homto C\otimes C$.}
  \label{fig:sd}
\end{figure}

We provide two examples in Figure~\ref{fig:sd}: we depict string diagrams within a box, with outer input ports listed above and outer output ports listed below; nodes are also represented with boxes, with inner output ports above and inner input ports below (note the reversal). The string diagram on the left is the identity on an object $A$, it has a single outer input port and a single outer output port, both labelled $A$, no node, and a single edge. The string diagram on the right has three outer input ports labelled $A,A,B$, three outer output ports each labelled $C$, and three nodes with various arities.

String diagrams can be composed vertically and horizontally. Vertical composition corresponds to categorical composition $({;})$: the outer output ports of the first diagram should match the outer input ports of the second one, which is plugged below it. Horizontal composition corresponds to tensor product $(\cdot)$: there are no constraints, the two diagrams are simply put next to each other, with their outer ports concatenated.

Using those operations, we may associate a string diagram to every morphism expression. In particular,
the diagram on the right of Figure~\ref{fig:sd} corresponds to the expression $(f\semi g)\cdot h$, with $f\colon A\otimes A\homto B$, $g\colon B\homto A$ and $h\colon B\homto C\otimes C$ (a leaf in the expression is mapped to the string diagram with exactly one node labelled by that leaf).

As announced before, the main benefit of string diagrams is that they abstract away irrelevant notational clutter. For instance, both sides of the bifunctoriality law
$(f\semi g) \cdot (f'\semi g') = (f\ccdot f')\ssemi(g\ccdot g')$ are mapped to the same diagram on the left below. Similarly, the three expression in the exchange law $I\ccdot j\ssemi i\ccdot J' = i\cdot j = i\ccdot J\ssemi I'\ccdot j$ correspond to the three pictures on the right, which are all equivalent up to continuous deformations in the plane.
\begin{align*}
  \vcenter{\hbox{\includegraphics[height=22mm]{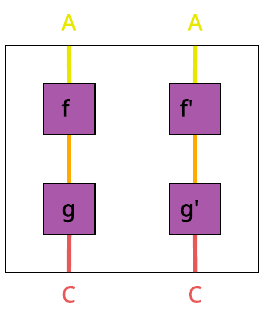}}}
  &&
  \vcenter{\hbox{\includegraphics[width=.2\linewidth,height=22mm]{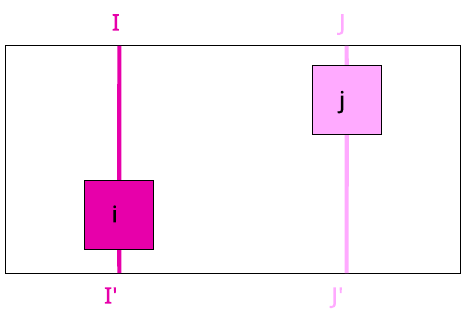}}}\quad
  \vcenter{\hbox{\includegraphics[width=.2\linewidth,height=22mm]{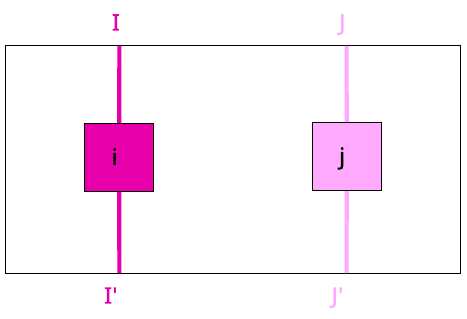}}}\quad
  \vcenter{\hbox{\includegraphics[width=.2\linewidth,height=22mm]{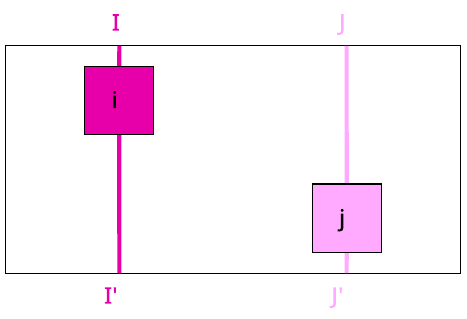}}}
\end{align*}
Similarly, the diagram on the right of Figure~\ref{fig:sd} also corresponds to the expressions $f\ccdot B \ssemi g \ccdot h$, $f\ccdot h \ssemi g \ccdot C \ccdot C$, or $f\ccdot B \ssemi B\ccdot h \ssemi g\ccdot C\ccdot C$, amongst others. In particular, this diagram makes it clear that $f$ and $h$ can be put next to each other, in case we would need to use an assumption of the shape $f\cdot h=\cdots$. This information was less obvious in the initial expression $(f\semi g)\cdot h$.

This approach is safe thanks to the following completeness theorem: all expressions denoting a given diagram are provably equal from the axioms of monoidal categories~\cite{MacLane:CWM}.

Note that string diagrams are inherently strict: they are indexed by lists of objects rather than trees of objects, and associators and unitors are all mapped to identity string diagrams. Nevertheless, the aforementioned completeness theorem holds for arbitrary monoidal categories: string diagrams abstract away both the bifunctoriality of the tensor product, and the coherence conditions from non-strict monoidal categories.

\section{Graphical interface}
\label{sec:gph}

We developed a standalone graphical editor for string diagrams.
This program is written in OCaml, using the \emph{vector graphics}~\cite{soft:vg} and \emph{store} libraries~\cite{ClementBasileMoineScherer:store}; it does not need to be trusted; we let it communicate with the proof assistant only via textual inputs and outputs.

This program is designed to perform three distinct tasks:
\begin{enumerate}
\item parse equational goals in monoidal categories; compute their string diagrams and render them graphically;
\item let the user delimit a subdiagram, detect whether this subdiagram matches an hypothesis of the goal, and substitute the subdiagram with the other member of the equation in that case;
\item extract expressions from string diagrams to translate graphical reasoning steps into textual ones.
\end{enumerate}
The first two items make it possible display the current state of a proof from the proof assistant, and to perform equational reasoning steps at the level of string diagrams. With appropriate library support (Section~\ref{sec:rocq}), the third one makes it possible to validate those steps in the proof assistant.

We discuss those three tasks in the next sections, illustrating them on the following running example.
Here we present this example in strict monoidal categories for the sake of clarity; we will see in Section~\ref{sec:rocq} how to formalise it in general ones, with no additional cost.

A \emph{monoid} in a strict monoidal category $\C$ is an object $M$ together with two morphisms $\mu\colon M\otimes M\homto M$ (the multiplication) and $\eta\colon 1\homto M$ (the unit), such that the following equations hold (associativity, left unitality, right unitality):
\begin{align}
  \label{ax:monoid}
  \tag{monoid axioms}
  \mu\cdot M \ssemi \mu &= M\cdot\mu \ssemi \mu &
  \eta\cdot M\ssemi \mu &= M &
   M\cdot\eta\ssemi \mu &= M
  \enspace.
\end{align}
Monoids in the monoidal category of endofunctors are just \emph{monads}, as known from the programming language community. A typical problem is to compose monads, and the standard solution consists in using distributive laws. In (strict) monoidal categories, this problem becomes: given two monoids $M,N$, define a monoid on $M\otimes N$.  A distributive law in that case is a morphism $x\colon N\otimes M\homto M\otimes N$ satisfying the following coherence laws:
\begin{align*}
  N\cdot \mu_M \ssemi x &= x\cdot M \ssemi M\cdot x \ssemi \mu_M\cdot N &%
  N\cdot \eta_M \ssemi x &= \eta_M\cdot N\\
  \mu_N\cdot M \ssemi x &= N\cdot x \ssemi x\cdot N \ssemi M\cdot \mu_N &%
  \eta_N\cdot M \ssemi x &= M\cdot\eta_N
  \enspace.
\end{align*}
The composite monoid can be defined as
\begin{align*}
  \mu_{M\otimes N}&\eqdef M\cdot x\cdot N \ssemi \mu_M\cdot \mu_N &
  \eta_{M\otimes N}&\eqdef \eta_M\cdot \eta_M\enspace.
\end{align*}
and we have to show the monoid axioms for it, assuming the monoid axioms for $M$ and $N$ and the above coherence axioms about the distributive law $x$. We focus on multiplication and the associativity axiom in the sequel; the same methodology applies to prove left and right unitality, as can be seen in the accompanying code.

\subsection{Parsing and rendering expressions}
\label{ssec:parse}

We use a text format which is flexible enough to encompass both Rocq goals under various notation systems, user annotations for standalone usage (e.g., about colours and shapes), and explicit diagram descriptions for debugging.
This format includes types and hypothesis declarations, as well as the current goal.
Parsing this format into abstract syntax trees is standard; we combine it with type inference to recover all implicit information and detect invalid inputs.

Formalising our monoid composition example results in the goal displayed in Figure~\ref{fig:rocqgoal},
where $m$ is the multiplication $\mu_M$, $n$ is $\mu_N$, and $mn$ is $\mu_{M\otimes N}$.
\begin{figure}
  \begin{minipage}{.36\linewidth}    
  \begin{lstlisting}
m : M⊗M ~> M
n : N⊗N ~> N
x : N⊗M ~> M⊗N
mA : m·M ; m ≡ M·m ; m
nA : n·N ; n ≡ N·n ; n
nx : n·M ; x ≡ N·x ; x·N ; M·n
mx : N·m ; x ≡ x·M ; M·x ; m·N
mn := M·x·N ; m·n
=============================
mn·M·N ; mn ≡ M·N·mn ; mn    
  \end{lstlisting}
  \end{minipage}\hfill
  \begin{minipage}{.4\linewidth}    
  \begin{lstlisting}
m : M⊗M ~> M
n : N⊗N ~> N
x : N⊗M ~> M⊗N
mA : m·M ; m ≡' M·m ; m
nA : n·N ; n ≡' N·n ; n
nx : n·M ; x ≡' N·x ;; x·N ;; M·n
mx : N·m ; x ≡' x·M ;; M·x ;; m·N
mn := M·x·N ;; m·n
================================
mn·M·N ;; mn ≡' M·N·mn ;; mn    
  \end{lstlisting}
  \end{minipage}
  \caption{Running example: composing the multiplications of two monoids $m$ and $n$, via a distributive law $x$. The goal on the right will be discussed only in Section~\ref{ssec:findmcl}.}
  \label{fig:rocqgoal}
\end{figure}

Most diagrams are initially described by terms, so that we can use the corresponding recursive structure to place them in the plane in a reasonable initial way. Then we use some force-directed dynamics in order to improve the placement: edges act as rubber bands, and vertical forces are adjusted so that nodes remain regularly distributed independently from the number of paths connecting them to the top and bottom interfaces. The initial placement is always a planar one; the user may freely move nodes but should be careful to do so in a continuous way, respecting faces---we plan to enforce this constraint in the future.

We assign colors to edges based on the name of the object labelling their endpoints. Similarly, we assign specific shapes to boxes of certain types: small triangles for morphisms of type $A\otimes A\homto A$ for some object $A$ (here, multiplications), small circles for morphisms of type $A\otimes B\homto B\otimes A$ (``crossings'', the distributive law here).

\begin{figure}
  \centering
  \includegraphics[scale=.3]{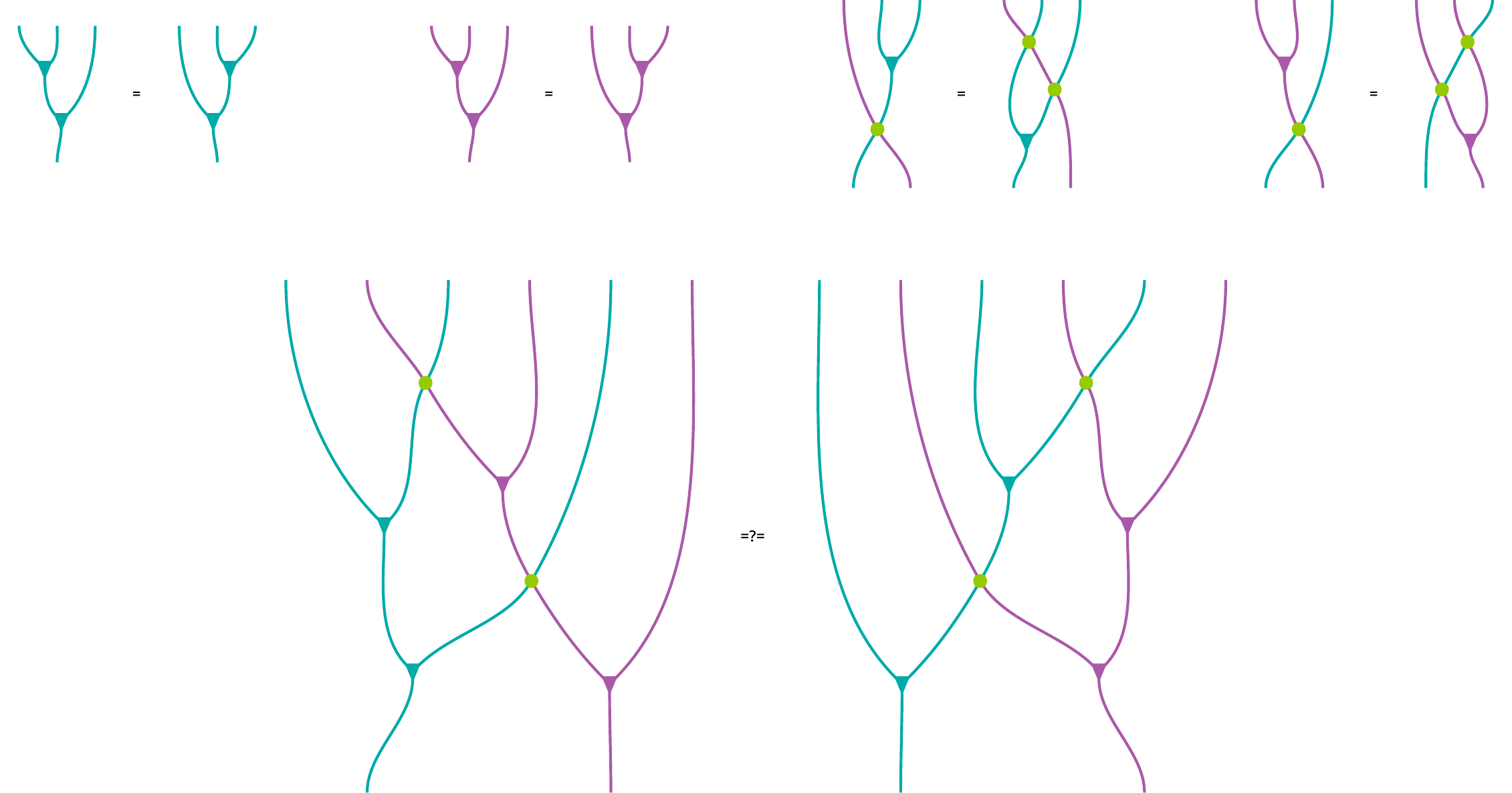}
  \caption{Rendering of the goal from Figure~\ref{fig:rocqgoal}.}
  \label{fig:sdgoal}
\end{figure}

When starting from the goal in Figure~\ref{fig:rocqgoal}, we obtain the rendering in Figure~\ref{fig:sdgoal}. Hypotheses are on the first row, and the conclusion is on the second one. We omit boxes and object names to alleviate the picture. 

\subsection{Isolating subdiagrams}
\label{ssec:isolate}

Thanks to this graphical rendering of the goal, it becomes clearer how to proceed. The first two associativity axioms make it possible to swap two consecutive triangles with the same colour, while the two coherence axioms make it possible for a triangle to traverse a crossing downwards, generating two new crossings. To equate the two diagrams on the bottom line, it suffices to let the two triangles in the middle traverse the two bottom crossings, and in parallel, to swap the two consecutive triangles on the left and on the right.

Our graphical interface makes it possible to proceed as follow: by surrounding a subdiagram with the pointer, we can create a box around it. If this subdiagram matches a member of one of the hypotheses then it is highlighted, and we can press a button to replace it with the other member of the hypothesis.
\begin{figure}
  \centering
  \begin{align*}
    \vcenter{\hbox{\includegraphics[width=.4\linewidth]{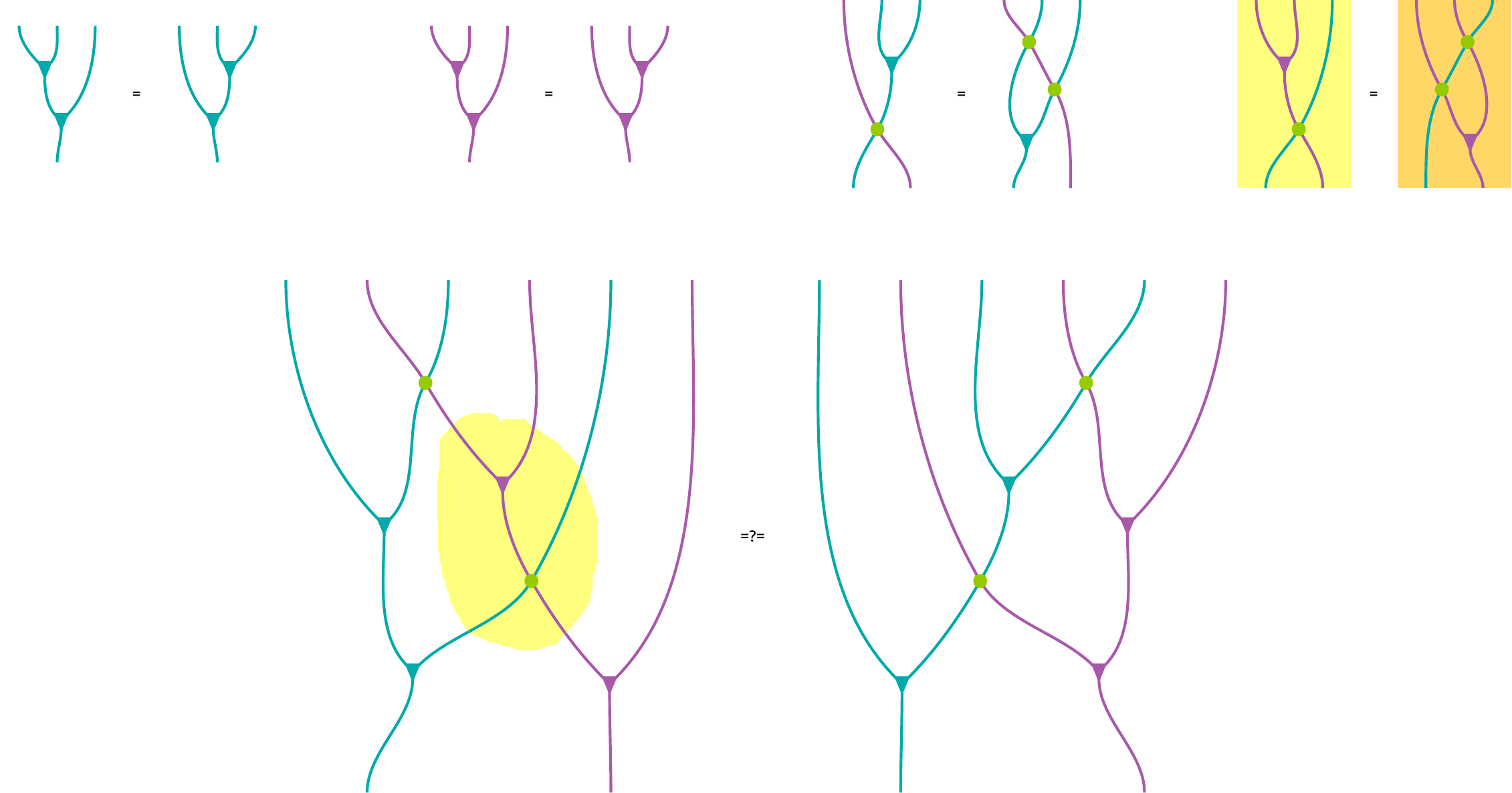}}}
    \qquad\vcenter{\hbox{$\Rightarrow$}}\qquad
    \vcenter{\hbox{\includegraphics[width=.4\linewidth]{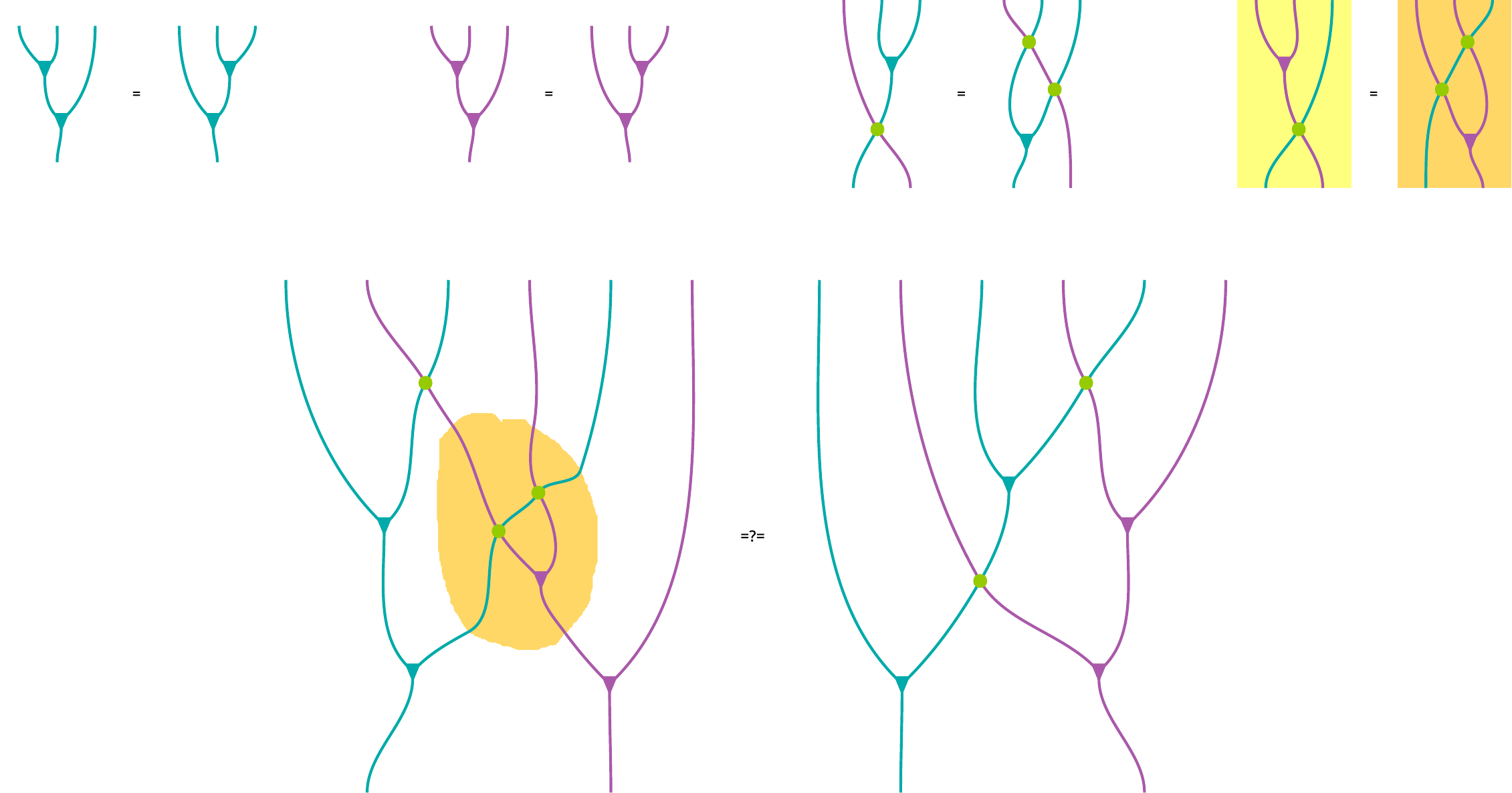}}}
  \end{align*}
  \caption{Performing a rewriting step graphically, by surrounding a matching subdiagram.}
  \label{fig:rwgoal}
\end{figure}

For this to work we need an algorithm to detect when two diagrams are equivalent, which is essentially an isomorphism test in our case, and a notion \emph{box}. In the implementation, this is done by generalising the notion of string diagram: we allow nodes to be labelled either by a morphism as before, or by a string diagram, recursively. In particular, the yellow and orange shapes in  Figure~\ref{fig:rwgoal} are new nodes in the top-level diagrams, with their own inner ports.

\noindent
\begin{minipage}{.8\linewidth}
Accordingly, our syntax for expressions includes a notation $[\_]$ for boxes.
For instance, for morphisms $f\colon A\otimes B\homto B$, $g\colon B\homto C$, and $h\colon B\otimes C\homto C$ of appropriate types, the diagram of the expression
$[f]\cdot B \ssemi [[B\cdot g] \ssemi h]$ is depicted on the right.
(While boxes created graphically by the user keep their hand-written shape, boxes arising from expressions are always rectangular.)  
\end{minipage}\hfill
\begin{minipage}{.15\linewidth}
  \includegraphics[width=.8\linewidth]{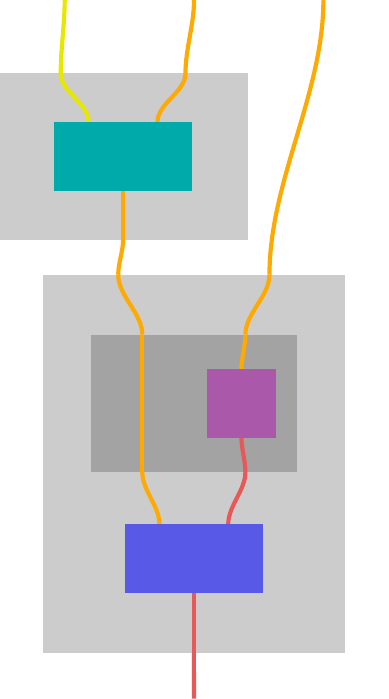}
\end{minipage}

\medskip
The two key operations on string diagrams with boxes are \emph{boxing} and \emph{unboxing}.
Unboxing inlines the content of a box within the surrounding diagram; it is mainly a matter of rewiring graphs. Boxing is more involved: given a polygon (the shape drawn by the user), we first have to compute the interface of the subdiagram to be created. To this end, we first compute the intersections of the polygon with the edges of the diagram and we check that each edge intersects at most twice; these intersections are computed in an oriented way, so that we can detect whether they correspond to inputs or to outputs. Then we check that there are not too many alternations between inputs and outputs along the polygon, so that we can extract a list of inputs and a list of outputs: the interface of our subdiagram. We finally check that all nodes within the polygon form a valid diagram with this interface, we turn it into a box node and we rewire the starting diagram accordingly.

\subsection{Extracting expressions}
\label{ssec:extract}

The aforementioned program should not be part of the trusted code-base. In order to validate the graphical rewriting steps in the proof assistant, we need a way to translate them back to the standard, textual format.

To this end, we need a function that extracts an expression from a given string diagram.
Potential many expressions yield a given string diagram, but any of them will do thanks to Rocq tactics we develop in Section~\ref{sec:rocq}.
Still, we define this function so that it returns human-friendly expressions.
In fact, it computes expressions which are in \emph{normal form}, in the following sense:
\begin{enumerate}
\item the expression is a composition of tensor products of identity morphisms and \emph{atoms}: boxes of normal forms or leaf morphisms;
\item each atom appears as right as possible in this composition.
\end{enumerate}
This corresponds to an algorithm where we read the string diagram line by line, bottom up, each time gathering as many nodes as possible, and recursively extracting expressions when we reach box nodes.

Consider for instance the string diagrams on the right of Figure~\ref{fig:sd}.
The first line we read, at the bottom, is $g\cdot h$; we could have read $g\cdot C\cdot C$ (or $C\cdot h$, or $C\cdot C\cdot C$), but these choices are not optimal.
Then the next and final line is $f\cdot B$, so that we return $f\cdot B \ssemi g\cdot h$.

Applying the same procedure to the bottom-left diagram of Figure~\ref{fig:sdgoal}, we get the expression
\begin{align*}
  M\cdot x\cdot N\cdot M\cdot N \ssemi M\cdot M\cdot n\cdot M\cdot N \ssemi m\cdot x\cdot N \ssemi m\cdot n\enspace.
\end{align*}
Note that is expression is quite different from the starting one ($mn\cdot M\cdot N \ssemi mn$ with $mn\eqdef M\cdot x\cdot N \ssemi m\cdot n$).

Applying it to the bottom-left diagram of Figure~\ref{fig:rwgoal}, we get the expression
\begin{align*}
  M\cdot x\cdot N\cdot M\cdot N \ssemi m\cdot [n\cdot M \ssemi x]\cdot N \ssemi m\cdot n\enspace.
\end{align*}
Observe here that the fact we have highlighted a subdiagram completely changes the extracted expression.
This is on purpose: now the left-hand side of hypothesis $nx$ matches the subexpression within brackets, and we get to see why we can rewrite using this hypothesis.

\section{Rocq library support}
\label{sec:rocq}

Now we describe the library support we provide in Rocq to formalise proofs in monoidal categories, potentially with the help of the previous graphical tool.

This involves two main tasks: setting up powerful notations that give the illusion of working in strict monoidal categories (Section~\ref{ssec:findmcl}), and designing a tactic to solve automatically any equation whose members denote the same string diagram. This second task requires us to prove a generalised version of MacLane's coherence theorem (Section~\ref{ssec:mcl}), and to implement a decision procedure for general expressions (Section~\ref{ssec:dec}).

\subsection{Inferring MacLane isomorphisms}
\label{ssec:findmcl}

If we try to generalise our definition of monoids in strict monoidal categories to the non-strict case, we see that the \ref{ax:monoid} do not type-check, as is. Indeed, we should insert appropriate associators and unitors
\begin{align*}
  \mu\cdot M \ssemi \mu &= \alpha_{M,M,M}\ssemi M\cdot\mu \ssemi \mu &
  \eta\cdot M\ssemi \mu &= \lambda_M & 
  M\cdot\eta\ssemi \mu &= \rho_M
  \enspace.
\end{align*}
The coherence axioms for distributive laws must be fixed in a similar way, as well as our definition of the multiplication and the unit of the composite monoid. We leave this as an exercise to the reader in the present paper: our goal in Rocq is to leave it to the assistant.

We achieve this task by designing a reflexive tactic~\cite{Boutin:reflexion} to infer automatically morphisms between two tensor-unit expressions denoting the same list of atomic objects: we say in this case that the tensor expressions are equivalent.
Thanks to this tactic, which we bind to the notation \lstinline{mcl}, we can write
\begin{center}
  \lstinline{mcl: A\otimes (B\otimes (1\otimes C))\_~>\_1\otimes (A\otimes B)\otimes C}
\end{center}
\lstinline{mcl} stands for ``MacLane isomorphism'', as it always infers (iso)morphisms built from  the associators $(\alpha)$ and the left and right unitors $(\lambda,\rho)$.

We use this notation to construct slightly more high-level ones\footnote{This is pseudocode, the actual notations are defined via appropriate definitions and a typeclass triggering our dedicated tactic.}:
\begin{center}
\lstinline{f ;; g := (f ; mcl ; g)        cast f := (mcl ; f ; mcl)        f ≡' g := (f ≡\_cast g)}
\end{center}
Intuitively: \lstinline{f;;g} makes it possible to compose two morphisms when the target of $f$ is only equivalent to the source of $g$, \lstinline{cast f} makes it possible to view $f$ as a morphism of a different yet equivalent type, and \lstinline{f ≡' g} makes it possible to compare morphisms with distinct yet equivalent types.

Using these notations, we can fix the goal on the left of Figure~\ref{fig:rocqgoal}, which does not typecheck, into the one on the right which does in any monoidal category: we can write expressions in monoidal categories as if they were definitionally strict.

The tactic works by reifying the tensor expressions into trees of objects, and showing that every tree of objects can be put in list normal form via a MacLane isomorphism. We call list of objects \emph{arities}; when the two trees denote the same arity, we get the desired morphism by going to this list normal form, and back. This is essentially how we get Lemma \lstinline{find_MacLane} in Figure~\ref{fig:maclane}, where we provide the key datatypes and functions. 

\begin{figure}
  \begin{lstlisting}
Inductive tree :=
| t_ob(A: 𝐂)
| t_unit
| t_tensor(a b: tree).
Fixpoint eval_tree: tree -> 𝐂 := ...

Inductive maclane: tree -> tree -> Type :=
| mcl_id: forall a, maclane a a
| mcl_comp: forall a b c, maclane a b -> maclane b c -> maclane a c
| mcl_tensor: forall a b c d, maclane a b -> maclane c d -> maclane (a⊗c) (b⊗d)
| mcl_inv: forall a b, maclane a b -> maclane b a
| mcl_assoc: forall a b c, maclane ((a⊗b)⊗c) (a⊗(b⊗c))
| mcl_unitl: forall a, maclane (1⊗a) a
| mcl_unitr: forall a, maclane (a⊗1) a.
Fixpoint eval_maclane a b (u: maclane a b): eval_tree a ~> eval_tree b := ...

Definition arity := list 𝐂.
Definition Nf: tree -> arity := ...

Lemma find_MacLane a b: Nf a = Nf b -> maclane a b.
Theorem MacLane a b (f g: maclane a b): eval_maclane f ≡ eval_maclane g.
  \end{lstlisting}
  \caption{Standard MacLane's coherence theorem.}
  \label{fig:maclane}
\end{figure}

\subsection{MacLane's coherence theorem}
\label{ssec:mcl}

The above approach is safe thanks to MacLane's coherence theorem for monoidal categories: all MacLane isomorphisms between two given tensor expressions are provably equal~\cite{MacLane:CWM}.
This property is crucial: it ensures that the statements we obtain do not depend on the specific choice of morphisms automatically inferred by our notations.

Accordingly, we prove this theorem, and we provide a high-level tactic \lstinline{MacLane}, that proves the equality of two morphism expressions differing only on MacLane isomorphisms. This tactic is a cornerstone for the subsequent development: we use it around 40 times explicitly to solve various goals when we verify the decision procedure from Section~\ref{ssec:dec}, and arount 170 times implicitly, via calls triggered by typeclass resolution.

We prove a first version of the theorem following closely the strategy proposed by Ilya Beylin and Peter Dybjer~\cite{BeylinD95}. Its formal statement is given in Figure~\ref{fig:maclane};
a technical point is that this coherence proof relies on a meta-coherence property: we need uniqueness of identity proofs on lists of objects of the starting category to conclude. For now we have added this requirement to our definition of monoidal category; we plan to lift it by using one more level of indirection during reification, so that we can rely on uniqueness of identity proofs on lists of natural numbers, which is derivable in CIC.

While this first version of MacLane's theorem should be enough for end-users of the library, we needed to refine it to formalise the proofs required in Section~\ref{ssec:dec} below.
Indeed, in addition to abstract objects and tensor trees of such objects, we frequently need to manipulate arities (lists of objects), to infer MacLane isomorphisms between types involving such arities, and to prove equality of such isomorphisms. For instance, if $\Gamma,\Delta$ are arities and \lstinline{ev: arity 𝐂 -> 𝐂} is a shorthand for the semantic evaluation function of arities (\lstinline{eval_arity}), we need the following automatic inferences:
\begin{center}
  \lstinline{mcl: ev ($\Gamma$++$\Delta$)\_~>\_ev $\Gamma$⊗ev $\Delta$}\qquad
  \lstinline{mcl: ev ($\Gamma$++(A⊗B)::$\Delta$)\_~>\_ev $\Gamma$⊗A⊗B⊗ev $\Delta$}
\end{center}
While the previous tools solve this issue in the case of concrete lists, we need a more general syntax (``generalised trees'') to deal with abstract ones. We manage to build on the standard MacLane's theorem in order to setup this infrastructure; still, the fact that we have to record trees in the type of MacLane isomorphisms makes it non-trivial from a technical perspective. The key definitions and results about this generalised version are given in Figure~\ref{fig:gmaclane}.

\begin{figure}
  \begin{lstlisting}
Inductive gtree :=
| gt_ob(A: 𝐂)
| gt_unit
| gt_tensor(s t: gtree)
| gt_trees(h: gtrees)
with gtrees :=
| gt_ar ($\Gamma$: arity)
| gt_nf (t: tree)    (* normal form of a plain tree *)
| gt_gnf (t: gtree)    (* normal form of a generalized tree *)
| gt_nil
| gt_cons (s: gtree) (h: gtrees)
| gt_app (h k: gtrees).
Fixpoint eval_gtree: gtree -> 𝐂 := ...
with eval_gtrees: gtrees -> arity := ...

Fixpoint gtree_tree: gtree -> tree := ...
Definition gmaclane (a b: gtree) := maclane (gtree_tree a) (gtree_tree b).
Definition eval_gmaclane a b (u: gmaclane a b): eval_gtree a ~> eval_gtree b
  := cast (eval_maclane u).

Definition gNf: gtree -> arity := ...

Lemma find_gMacLane a b: gNf a = gNf b -> gmaclane a b.
Theorem gMacLane a b (f g: gmaclane a b): eval_gmaclane f ≡ eval_gmaclane g.
  \end{lstlisting}
  \caption{Extended MacLane's coherence theorem.}
  \label{fig:gmaclane}
\end{figure}

\subsection{Decision procedure}
\label{ssec:dec}

The aforementioned tactic \lstinline{MacLane} makes it possible to deal with the bureaucracy induced by MacLane morphisms: it automates the use of the \ref{ax:triangle} and \ref{ax:pentagon} axioms from the definition of monoidal category, together with naturality of the associators and unitors.
It remains to help the user with bifunctoriality, which is also extremely painful to use manually.

To this end, we develop a tactic \lstinline{mcat} to solve equations in monoidal categories.
Given two morphism expressions $f,g$, this tactic proves $f=g$ when $f$ and $g$ denote the same string diagram. However, formalising string diagrams and equivalence up to continuous deformations seems quite challenging. Instead, we choose to restrict to the case where all nodes have at least one output port (i.e., there is no morphism of type $A\homto 1$ for some $A$).
This restriction makes it possible to completely bypass this difficulty.
Indeed, when all nodes have at least one output port, all nodes in a string diagrams are reachable bottom-up from its outputs, and the expressions in normal form computed by our standalone graphical editor are unique (see Appendix~\ref{app:emptytarget} for a counter-example in the presence of nodes with empty target). Therefore, to solve this fragment, it suffices to show that every morphism expression can be put in normal form using the axioms of monoidal categories.

Even if this process is simpler than formalising string diagrams, it remains non-trivial.
We proceed in two steps. First we define a syntax for end-user morphisms indexed by (generalised) trees, which we translate to a syntax of strict morphisms, indexed by arities. The datatypes and the key definitions and statements are given in Figure~\ref{fig:strictify}; in a sense this step corresponds to strictification, at a reified level: the source expressions contain non-trivial MacLane morphisms (constructor \lstinline{t_mcl}), while the target expressions restrict this construction to plain identity morphisms (constructor \lstinline{a_id}). Theorem \lstinline{eval_gNf_mor} can be proved in a few lines including six calls to the aforementioned \lstinline{(g)MacLane} tactic.
\begin{figure}
  \begin{lstlisting}
Inductive tmor: gtree -> gtree -> Type :=
| t_var: forall a b, (eval_gtree a ~> eval_gtree b) -> tmor a b
| t_mcl: forall a b, gNf a = gNf b -> tmor a b
| t_comp: forall a b c, tmor a b -> tmor b c -> tmor a c
| t_tens: forall a b c d, tmor a b -> tmor c d -> tmor (gt_tens a c) (gt_tens b d).
Fixpoint eval_tmor a b (u: tmor a b): eval_gtree a ~> eval_gtree b := ...

Inductive amor: arity -> arity -> Type :=
| a_var: forall n m, (eval_arity n ~> eval_arity m) -> amor n m
| a_id: forall n, amor n n
| a_comp: forall n m p, amor n m -> amor m p -> amor n p
| a_tens: forall n m p q, amor n m -> amor p q -> amor (n++p) (m++q).
Fixpoint eval_amor n m (u: amor n m): eval_arity n ~> eval_arity m := ...

Fixpoint gNf_mor a b (u: tmor a b): amor (gNf a) (gNf b) := ...
Theorem eval_gNf_mor a b (f: tmor a b): eval_tmor f ≡' eval_amor (gNf_mor f). 
  \end{lstlisting}  
  \caption{Strictifying reified morphism expressions.}
  \label{fig:strictify}
\end{figure}

This first step is important: it makes it possible to work on a simplified syntax for the subsequent normalisation function. Recall that a normal form is a composition of tensor products, where atoms in the tensor products are pushed as far as possible to the right (i.e., towards bottom in the string diagrams). We use the datatype given in Figure~\ref{fig:normalforms}: a normal form is a list of rows, where a row is a list of atoms interleaved with identity morphisms. This datatype does not ensure that atoms are pushed as far as possible to the right (i.e., towards the tail of the list of rows), and it is our responsibility when we define the normalisation function to enforce this requirement.

\begin{figure}
  \begin{lstlisting}
Inductive row: arity -> arity -> Type :=
| rnil: forall n, row n n
| rcons: forall n m m' (u: eval_arity m ~> eval_arity m') p p',
         row p p' -> row (n ++ m ++ p) (n ++ m' ++ p').
Fixpoint eval_row n m (r: row n m): eval_arity n ~> eval_arity m := ...

Inductive nmor n: arity -> Type :=
| nnil: nmor n n
| ncons: forall m p, nmor n m -> row m p -> nmor n p.
Fixpoint eval_nmor n m (u: nmor n m): eval_arity n ~> eval_arity m := ...

Fixpoint norm n m (u: amor n m): nmor n m := ...
Theorem eval_norm n m (u: amor n m): eval_nmor (norm u) ≡ eval_amor u.
  \end{lstlisting}
  \caption{Datatype for normal forms, and normalisation function.}
  \label{fig:normalforms}
\end{figure}

\begin{figure}
  \includegraphics[width=.45\linewidth]{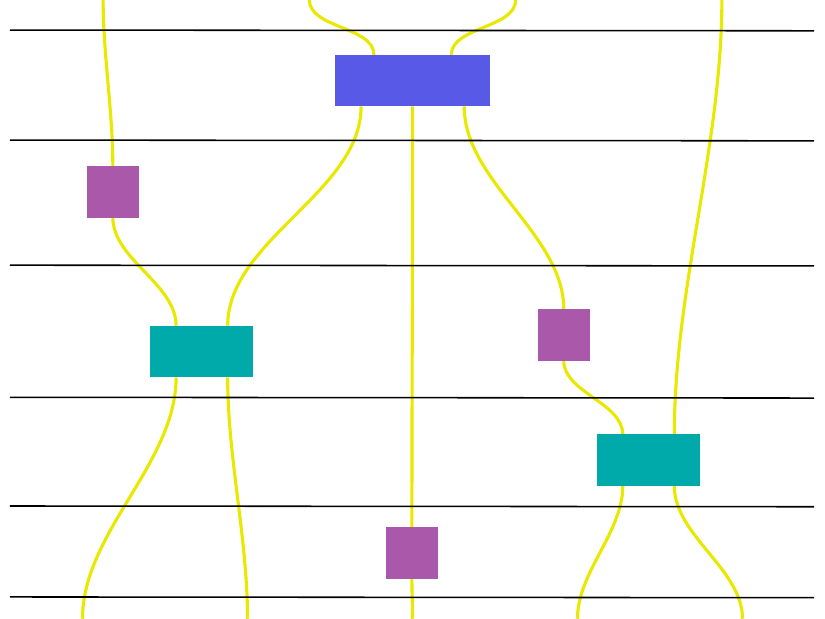}\hfill
  \includegraphics[width=.45\linewidth]{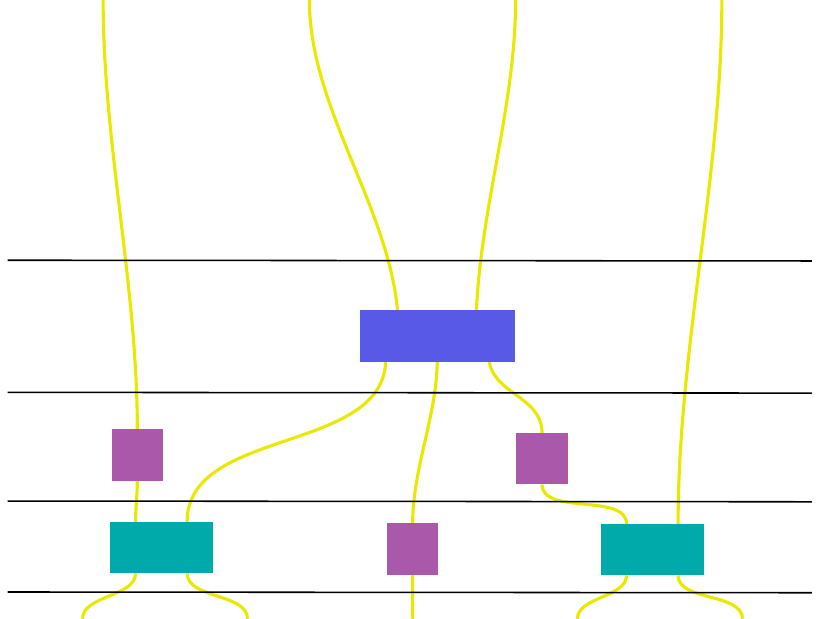}
  \caption{Playing tetris on string diagrams: the five rows on the left have to be normalised into the three rows on the right.}
  \label{fig:tetris}
\end{figure}

The normalisation rests on several auxiliary functions (13).
A first part of it consists in translating general morphisms into lists of rows, which is not too hard.
The rest of it intuitively implements the tetris game: given a list of rows, let the atoms jump from one row to the next one down the list, as far as possible, as illustrated in Figure~\ref{fig:tetris}.

This problem reduces to defining a function that takes two rows, and returns either a single row if all atoms from the first one can be inserted in the second one, or the two rows where all atoms from the first one that can be inserted in the second one have been so. Writing such a function is an interesting exercise \emph{per se}; here it is complicated  by the fact that rows are dependently typed: we should carry all case analyses in such a way that we preserve type annotations, and that we do so in a way that does not block computations later when we execute the normalisation function.

We implement this function using 9 additional functions working on \emph{relative arities}, that make it possible to compare two rows whose internal interfaces are slightly desynchronised (by a relative arity), due to some blocking atom. The final function (\lstinline{zrcomp}) is written in Ltac, it consists of 478 words spanning over 63 lines.
The correctness of our normalisation function (Theorem \lstinline{eval_norm}) spans over 400 lines, 280 of which being dedicated to the analysis of the function \lstinline{zrcomp}.
These line numbers remain reasonable because a lot of the monoidal reasoning can be performed via implicit calls to the \lstinline{MacLane} tactic, which is triggered by typeclass resolution when we apply various simplification lemmas. 

We finally implement reification (in Ltac for now), and we pack everything into a reflexive tactic \lstinline{mtac}. For instance, we may use it to prove that any two expressions for the string diagram on the right of Figure~\ref{fig:sd} are equal:
\begin{lstlisting}
Goal forall f: A⊗A ~> B, forall g: B ~> C, forall h: B ~> C⊗C, (f;g)·h ≡ f·h ; g·C·C.
Proof. intros. mcat. Qed.
\end{lstlisting}

\subsection{Composing monoids, formally}
\label{ssec:wrap}

Now we have all the necessary material to complete a formal proof of our running example about composition of monoids.
Starting from the Rocq goal in Figure~\ref{fig:rocqgoal}, we can render it graphically with the graphical editor (Figure~\ref{fig:sdgoal}) and proceed with the four graphical rewriting steps (a first one being depicted in Figure~\ref{fig:rwgoal}).
Then the editor detects that we are done, and we may ask it to export a proof script which we can copy-paste and execute in Rocq.
This proof is given in Figure~\ref{fig:finalproof:auto}; it consists of four transitivity steps which make explicit the pattern which is next rewritten, within brackets. Each of these transitivity step is validated by a call to \lstinline{mcat}: they consist in moving from one expression for some diagram to another expression for the same diagram. A final call to \lstinline{mcat} validates the fact that we have indeed reached the same diagram after the four rewriting steps. While boxes have a specific behaviour in the diagram editor (Section~\ref{ssec:isolate}), they are just implemented via an identity function in Rocq, bound to the bracket notation to highlight the pattern to be rewritten.

\begin{figure}
  \begin{lstlisting}
Lemma mnA: mn·M·N ;; mn ≡' M·N·mn ;; mn.
Proof.
  unfold mn.

  transitivity (M·x·N·M·N ;; m·[n·M ; x]·N ;; m·n). mcat.
  rewrite nx.

  transitivity (M·N·M·x·N ;; M·[N·m ; x]·n ;; m·n). 2: mcat.
  rewrite mx.

  transitivity (M·x·x·N ;; M·M·x·N·N ;; M·M·M·n·N ;; [m·M ; m]·n). mcat.
  rewrite mA.
  
  transitivity (M·x·x·N ;; M·M·x·N·N ;; M·m·N·N·N ;; m·[N·n ; n]). 2: mcat.
  rewrite -nA.
  
  mcat.
Qed.    
  \end{lstlisting}
  \caption{Associativity of the multiplication for a composite monoid, formally.}
  \label{fig:finalproof:auto}
\end{figure}

A key feature of the proof we obtained is that it is readable, robust to slight changes in the statement, and easily editable.
Moreover, it can be replayed step by step using the graphical editor: each intermediate goal can be imported back in the editor to see what are the diagrams at that point.
We may also optimise the proof manually: some rewriting steps can be performed in parallel, and we can use the diagram editor to extract expressions on which two rewriting steps can be applied, simply by surrounding the two matches and asking the editor to export the corresponding expressions. This is how we obtained the variant in Figure~\ref{fig:finalproof:opt}, for instance.

\begin{figure}
  \begin{lstlisting}
Lemma mnA: mn·M·N ;; mn ≡' M·N·mn ;; mn.
Proof.
  unfold mn.

  transitivity (M·x·N·M·N ;; M·M·[n·M ; x]·N ;; [m·M ; m]·n). mcat.
  rewrite nx.
  rewrite mA.

  transitivity (M·N·M·x·N ;; M·[N·m ; x]·N·N ;; m·[N·n ; n]). 2: mcat.
  rewrite mx.
  rewrite -nA.

  mcat.
Qed.    
  \end{lstlisting}
  \caption{Manually optimised formal proof.}
  \label{fig:finalproof:opt}
\end{figure}

For the sake of comparison, we provide an alternative ``paper proof'' via commutative diagrams in Appendix~\ref{app:explicit}.
The complete example, where we also deal with the unit laws of the composite monoid, can be found in the accompanying code, as well as the exercise consisting in composing three monoids, using three distributive laws.

\section{Related work}
\label{sec:rw}

There are several standalone domain-specific proof assistants or visualisers for various kinds of categories in the literature:
Globular~\cite{bar_et_al:LIPIcs.FSCD.2016.34}, homotopy.io~\cite{corbyn_et_al:LIPIcs.FSCD.2024.30}, 
Quantomatic~\cite{Quantomatic25}, or Chyp~\cite{soft:chyp} which is a graphical prover for symmetric monoidal categories, implementing the theory from~\cite{Bonchi_Gadducci_Kissinger_Sobocinski_Zanasi_2022}. The latter is the closest in spirit to our work, but none of them connects to general purpose proof assistants.

Sam Ezeh proposed a visualiser for string diagrams in Lean~4~\cite{ezeh:LIPIcs.ITP.2024.41}, and makes it possible to perform rewriting steps from the visualiser, but only in the strict monoidal category of endofunctors, only for specific forms of hypotheses (naturality, associativity of a monad), and only modulo associativity of composition.

In Rocq, Bhakti Shash et al. developed a visualiser for symmetric monoidal categories~\cite{Shah_2025}, together with applications to the ZX calculus~\cite{lehmann2026vyzx}.
They formalise MacLane's coherence theorem, but they do not provide tactics to infer MacLane isomorphisms, nor a way to perform rewriting steps graphically.

Leaving monoidal categories aside, Ambroise Lafont designed a diagram editor~\cite{lafontDiagramEditorMechanise2024} that makes it possible to perform categorical proofs graphically and export them to Rocq. 
In the same vein, Luc Chabassier designed an integrated graphical interface for Rocq with similar objectives~\cite{chabassierGraphicalInterfaceCategory2025,chabassierAspectsCategoryTheory2025}.
In both cases, the tools do not deal with string diagrams but with standard commutative diagrams in general category theory (as in the explicit proof given in Appendix~\ref{app:explicit}).

\section{Directions for future work}
\label{sec:fw}

Besides improving \lstinline{mcat} to make it complete even in the presence of empty target morphisms, which seems quite challenging, let us mention three main directions for future work.

First we would like to extend the Rocq development to deal with \emph{bicategories}; those generalise monoidal categories (these are one-object bicategories), and they share the same string diagrams. Extending the graphical editor would be easy, but extending the Rocq decision tactic would require to add one more layer of dependent types to our normalisation procedure, which is likely to be unpleasant.

Second we would like to deal with \emph{symmetric} monoidal categories. On the one hand, the diagrams become simpler: the planarity constraint disappears, and diagrams become plain graphs: ``only connectivity matters''. On the other hand, it is more difficult to define a satisfactory notion normal form for expressions, as we have to record permutations at several places in expressions.

Lastly, while we advocate for a clear separation between the graphical editor and the proof assistant, we could also propose a more integrated version. This would make it possible to render the current goal without any user interaction, but also to design tactics that do not display any string diagram but compute them in the background to implement specific tasks, for instance to find matches modulo the monoidal category axioms automatically, or to implement specific equality saturation procedures.

\appendix

\section{Counter-example to uniqueness of normal forms}
\label{app:emptytarget}

In the presence of nodes with empty targets, the tactic \lstinline{mcat} is not longer complete.
This is because in that case, the normal forms we compute are no longer unique.

\begin{figure}
  \begin{align*}
    \includegraphics[width=.11\linewidth]{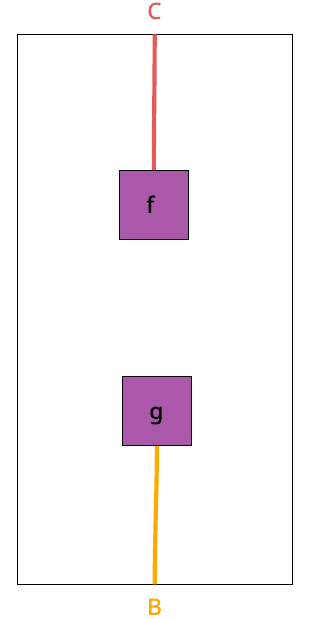}&&
    \includegraphics[width=.1\linewidth]{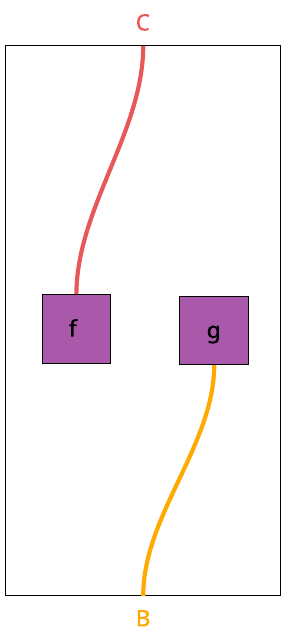}&&
    \includegraphics[width=.1\linewidth]{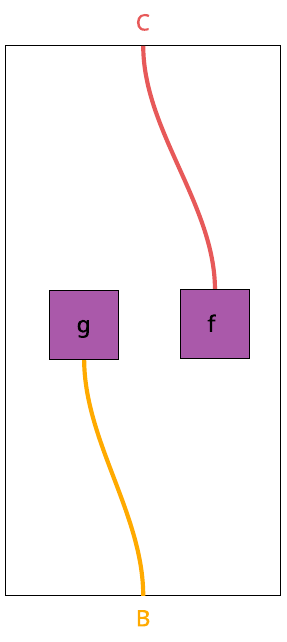}
  \end{align*}  
  \caption{Non uniqueness of normal forms in the presence of morphisms without outputs.}
  \label{fig:emptytgt}
\end{figure}

Consider for instance morphisms $f\colon C\homto 1$ and $g\colon 1 \homto B$. The diagram for $f;g$ is depicted below on the left of Figure~\ref{fig:emptytgt}, and it is equivalent to the two diagrams on the right: when playing the tetris game, we may chose the move $f$ either on the left, or on the right of $g$, resulting in two distinct normal forms: $f\cdot g$ and $g\cdot f$.

\section{An explicit proof for the associativity of the composite multiplication}
\label{app:explicit}

Here we provide an explicit proof for the associativity of the composite multiplication $mn$ (Figure~\ref{fig:rocqgoal}).
This proof is displayed as commutative diagram (not a string diagram) in Figure~\ref{fig:altproof};
the two squares marked with uni-coloured disks correspond to the two steps where with use associativity of the starting multiplications; the two pentagons marked with mixed-colour disks correspond to the two steps where we use the coherence axioms of the distributive law. The remaining 8 squares are bifunctoriality, and the four triangles are the exchange law. Those 12 steps are entirely implicit in the string diagrammatic proofs we formalised (Figures~\ref{fig:finalproof:auto} and \ref{fig:finalproof:opt}): they are covered automatically by the \lstinline{mcat} tactic.

\begin{figure}
\definecolor{monadm}{rgb}{0.0, 0.4, 0.4}
\definecolor{monadn}{rgb}{0.4, 0.1, 0.4}
\definecolor{monadmn}{rgb}{1.0, 0.4, 0.1}
\definecolor{monadx}{rgb}{0.3, 0.6, 0.0}
\newcommand\M{{\color{monadm}M}}
\newcommand\muM{{\color{monadm}m}}
\newcommand\N{{\color{monadn}N}}
\newcommand\muN{{\color{monadn}n}}
\newcommand\MN{{\color{monadmn}M}}
\newcommand\lam{{\color{monadx}x}}
\newcommand\mmu{\mu\!\!\!\mu}
\newcommand\yinyang[3][!70!white]{%
  \tikz{
    \begin{scope}
      \clip (0,0) circle (.25cm);
      \fill[#2#1] (0cm,.25cm) rectangle (-.25cm, -.25cm);
    \end{scope}
    \begin{scope}
      \clip (0,0) circle (.25cm);
      \fill[#3#1] (0cm,.25cm) rectangle (.25cm, -.25cm);
    \end{scope}
    \fill[#2#1] (0,0.125) circle (0.125cm);
    \fill[#3#1] (0,-0.125) circle (0.125cm);
  }
}
\begin{align*}
  \xymatrix @C=.3em @R=2em {
  (\M\N)^3
    \ar[rr]^{\M\N\M\lam\N}
    \ar[d]_{\M\lam\N\M\N}
  &&\M\N\M^2\N^2
     \ar[rr]^{\M\N\M^2\muN}
     \ar[d]^{\M\lam\M\N^2}
  \ar@/^3em/[rrrr]^{\M\N\muM\muN}
  &&\M\N\M^2\M
     \ar[rr]^{\M\N\muM\N}
     \ar[d]^{\M\lam\M\N}
      \ar@[white][rrdd]|{\yinyang{monadm}{monadx}}
  &&(\M\N)^2
     \ar[dd]^{\M\lam\N}
  \\
  \M^2\N^2\M\N
  \ar[rr]^{\M^2\N\lam\N}
  \ar[dd]_{\muM\N^2\M\N}
  \ar@/_3em/[dddd]_{\muM\muN\M\N}
  &&\M^2\N\M\N^2
    \ar[rr]^{\M^2\N\M\muN}
    \ar[ld]_{\muM\N\M\N^2}
    \ar[rd]^{\M^2\lam\N^2}       
  &&\M^2\N\M\N
    \ar[d]^{\M^2\lam\N}
  \\
  &\M\N\M\N^2
    \ar[dr]_{\M\lam\N^2}
  &&\M^3\N^3
    \ar[r]^{\M^3\N\muN}
    \ar[dl]_{\muM\M\N^3}
    \ar[dd]_{\M^3\muN\N}
     \ar@[white][rdd]|{\yinyang{monadn}{monadn}}
  &\M^3\N^2
    \ar[rr]^{\M\muM\N^2}
    \ar[dr]^{\muM\M\N^2}
    \ar[dd]^{\M^3\muN}
      \ar@[white][rrd]|{\hspace{2em}\yinyang{monadm}{monadm}}
  &&\M^2\N^2
     \ar[d]_{\muM\N^2}
     \ar@/^3em/[ddd]^{\muM\muN}
  \\
  \M\N^2\M\N
  \ar[ru]_{\M\N\lam\N}
  \ar[dd]^{\M\muN\M\N}
  &&\M^2\N^3
     \ar[dd]_{\M^2\muN\N}
  &&&\M^2\N^2
      \ar[r]_{\muM\N^2}
      \ar[dd]_{\M^2\muN}
  &\M\N^2
    \ar[dd]_{\M\muN}
  \\
  &&&\M^3\N^2
      \ar[ld]^{\muM\M\N^2}
      \ar[r]_{\M^3\muN}
  &\M^3\N
      \ar[rd]_{\muM\M\N}
  \\
  (\M\N)^2
  \ar[rr]_{\M\lam\N}
    \ar@[white][rruu]|{\yinyang{monadn}{monadx}} 
  &&\M^2\N^2
  \ar[rrr]_{\M^2\muN}
     \ar@/_3em/[rrrr]_{\muM\muN}
  &&&\M^2\N
  \ar[r]_{\muM\N}
  &\M\N\\    
  }
\end{align*}
  \caption{Proof for the associativity of the composite monoid multiplication, using commutative diagrams rather than string diagrams.}
  \label{fig:altproof}
\end{figure}

Also note that the commutative diagram proof in Figure~\ref{fig:altproof} is only a proof in \emph{strict} monoidal categories. A proof in general monoidal categories would not fit in a page as it would require the insertion of many associators and unitors, as well as many uses of the \ref{ax:triangle} and \ref{ax:pentagon} coherence axioms (which are also dealt with by \lstinline{mcat} in our case, which supersedes \lstinline{MacLane}).

\bibliography{pous,main}

\begin{thebibliography}{10}

\bibitem{Ahrensetal2021}
Benedikt Ahrens, Dan Frumin, Marco Maggesi, Niccolò Veltri, and Niels van~der
  Weide.
\newblock Bicategories in univalent foundations.
\newblock {\em Mathematical Structures in Computer Science},
  31(10):1232–1269, 2021.
\newblock \href {https://doi.org/10.1017/S0960129522000032}
  {\path{doi:10.1017/S0960129522000032}}.

\bibitem{ClementBasileMoineScherer:store}
Cl\'{e}ment Allain, Basile Cl\'{e}ment, Alexandre Moine, and Gabriel Scherer.
\newblock Snapshottable stores.
\newblock {\em Proc. ACM Program. Lang.}, 8(ICFP), August 2024.
\newblock \href {https://doi.org/10.1145/3674637} {\path{doi:10.1145/3674637}}.

\bibitem{ArsacHP26}
Samuel Arsac, Russ Harmer, and Damien Pous.
\newblock Adhesive category theory for graph rewriting in rocq.
\newblock In Kathrin Stark, Yannick Zakowski, Nikhil Swamy, and Nicolas
  Tabareau, editors, {\em Proceedings of the 15th {ACM} {SIGPLAN} International
  Conference on Certified Programs and Proofs, {CPP} 2026, Rennes, France,
  January 12-13, 2026}, pages 59--74. {ACM}, 2026.
\newblock \href {https://doi.org/10.1145/3779031.3779105}
  {\path{doi:10.1145/3779031.3779105}}.

\bibitem{bar_et_al:LIPIcs.FSCD.2016.34}
Krzysztof Bar, Aleks Kissinger, and Jamie Vicary.
\newblock {Globular: An Online Proof Assistant for Higher-Dimensional
  Rewriting}.
\newblock In Delia Kesner and Brigitte Pientka, editors, {\em 1st International
  Conference on Formal Structures for Computation and Deduction (FSCD 2016)},
  volume~52 of {\em Leibniz International Proceedings in Informatics (LIPIcs)},
  pages 34:1--34:11, Dagstuhl, Germany, 2016. Schloss Dagstuhl --
  Leibniz-Zentrum f{\"u}r Informatik.
\newblock \url{http://globular.science}.
\newblock \href {https://doi.org/10.4230/LIPIcs.FSCD.2016.34}
  {\path{doi:10.4230/LIPIcs.FSCD.2016.34}}.

\bibitem{BeylinD95}
Ilya Beylin and Peter Dybjer.
\newblock Extracting a proof of coherence for monoidal categories from a proof
  of normalization for monoids.
\newblock In Stefano Berardi and Mario Coppo, editors, {\em Types for Proofs
  and Programs, International Workshop TYPES'95, Torino, Italy, June 5-8, 1995,
  Selected Papers}, volume 1158 of {\em Lecture Notes in Computer Science},
  pages 47--61. Springer, 1995.
\newblock \href {https://doi.org/10.1007/3-540-61780-9\_61}
  {\path{doi:10.1007/3-540-61780-9\_61}}.

\bibitem{Bonchi_Gadducci_Kissinger_Sobocinski_Zanasi_2022}
Filippo Bonchi, Fabio Gadducci, Aleks Kissinger, Pawel Sobocinski, and Fabio
  Zanasi.
\newblock String diagram rewrite theory ii: Rewriting with symmetric monoidal
  structure.
\newblock {\em Mathematical Structures in Computer Science}, 32(4):511–541,
  2022.
\newblock \href {https://doi.org/10.1017/S0960129522000317}
  {\path{doi:10.1017/S0960129522000317}}.

\bibitem{borceuxHandbookCategoricalAlgebra1994}
Francis Borceux.
\newblock {\em Handbook of {{Categorical Algebra}}: {{Volume}} 1: {{Basic
  Category Theory}}}, volume~1 of {\em Encyclopedia of {{Mathematics}} and Its
  {{Applications}}}.
\newblock Cambridge University Press, 1994.
\newblock \href {https://doi.org/10.1017/CBO9780511525858}
  {\path{doi:10.1017/CBO9780511525858}}.

\bibitem{Boutin:reflexion}
Samuel Boutin.
\newblock Using reflection to build efficient and certified decision
  procedures.
\newblock In {\em Proceedings of the Third International Symposium on
  Theoretical Aspects of Computer Software}, TACS '97, page 515–529, Berlin,
  Heidelberg, 1997. Springer-Verlag.
\newblock \href {https://doi.org/10.5555/645869.668533}
  {\path{doi:10.5555/645869.668533}}.

\bibitem{Benabou63}
Jean Bénabou.
\newblock Catégories avec multiplication.
\newblock {\em Compte rendus de l'académie des sciences}, 256, 1963.

\bibitem{Benabou72}
Jean Bénabou.
\newblock Les catégories multiplicatives.
\newblock {\em Séminaire de mathématique pure}, 27, 1972.
\newblock URL:
  \url{https://ncatlab.org/nlab/files/Benabou-CategoriesMultiplicatives.pdf}.

\bibitem{soft:vg}
Daniel Bünzli.
\newblock Vector graphic, 2013.
\newblock URL: \url{https://erratique.ch/software/vg}.

\bibitem{chabassierAspectsCategoryTheory2025}
Luc Chabassier.
\newblock {\em Aspects of Category Theory in Proof Assistants}.
\newblock Thèse de doctorat, Universit\'e Paris-Saclay, July 2025.
\newblock URL: \url{https://theses.fr/2025UPASG055}.

\bibitem{chabassierGraphicalInterfaceCategory2025}
Luc Chabassier.
\newblock A {{Graphical Interface}} for {{Category Theory Proofs}} in {{Coq}}.
\newblock {\em Electronic Proceedings in Theoretical Computer Science},
  419:28--41, May 2025.
\newblock URL: \url{http://arxiv.org/abs/2505.13473}, \href
  {https://arxiv.org/abs/2505.13473} {\path{arXiv:2505.13473}}, \href
  {https://doi.org/10.4204/EPTCS.419.2} {\path{doi:10.4204/EPTCS.419.2}}.

\bibitem{cohenHierarchyBuilderAlgebraic2020}
Cyril Cohen, Kazuhiko Sakaguchi, and Enrico Tassi.
\newblock Hierarchy {{Builder}}: Algebraic hierarchies made easy in {{Coq}}
  with {{Elpi}}.
\newblock In {\em {{FSCD}} 2020 - 5th {{International Conference}} on {{Formal
  Structures}} for {{Computation}} and {{Deduction}}}, volume 167 of {\em
  Leibniz {{International Proceedings}} in {{Informatics}} ({{LIPIcs}})}, pages
  34:1 -- 34:21. Schloss Dagstuhl -- Leibniz-Zentrum f{\"u}r Informatik, June
  2020.
\newblock \url{https://github.com/math-comp/hierarchy-builder}.
\newblock \href {https://doi.org/10.4230/LIPIcs.FSCD.2020.34}
  {\path{doi:10.4230/LIPIcs.FSCD.2020.34}}.

\bibitem{corbyn_et_al:LIPIcs.FSCD.2024.30}
Nathan Corbyn, Lukas Heidemann, Nick Hu, Chiara Sarti, Calin Tataru, and Jamie
  Vicary.
\newblock {homotopy.io: A Proof Assistant for Finitely-Presented Globular
  n-Categories}.
\newblock In Jakob Rehof, editor, {\em 9th International Conference on Formal
  Structures for Computation and Deduction (FSCD 2024)}, volume 299 of {\em
  Leibniz International Proceedings in Informatics (LIPIcs)}, pages
  30:1--30:26, Dagstuhl, Germany, 2024. Schloss Dagstuhl -- Leibniz-Zentrum
  f{\"u}r Informatik.
\newblock \url{https://beta.homotopy.io}.
\newblock \href {https://doi.org/10.4230/LIPIcs.FSCD.2024.30}
  {\path{doi:10.4230/LIPIcs.FSCD.2024.30}}.

\bibitem{ezeh:LIPIcs.ITP.2024.41}
Sam Ezeh.
\newblock {Graphical Rewriting for Diagrammatic Reasoning in Monoidal
  Categories in Lean4}.
\newblock In Yves Bertot, Temur Kutsia, and Michael Norrish, editors, {\em 15th
  International Conference on Interactive Theorem Proving (ITP 2024)}, volume
  309 of {\em Leibniz International Proceedings in Informatics (LIPIcs)}, pages
  41:1--41:8, Dagstuhl, Germany, 2024. Schloss Dagstuhl -- Leibniz-Zentrum
  f{\"u}r Informatik.
\newblock \href {https://doi.org/10.4230/LIPIcs.ITP.2024.41}
  {\path{doi:10.4230/LIPIcs.ITP.2024.41}}.

\bibitem{garillotPackagingMathematicalStructures2009}
Fran{\c c}ois Garillot, Georges Gonthier, Assia Mahboubi, and Laurence Rideau.
\newblock Packaging {{Mathematical Structures}}.
\newblock In Stefan Berghofer, Tobias Nipkow, Christian Urban, and Makarius
  Wenzel, editors, {\em Theorem Proving in Higher Order Logics. TPHOLs 2009},
  LNCS, pages 327--342. Springer, Berlin, Heidelberg, 2009.
\newblock \href {https://doi.org/10.1007/978-3-642-03359-9_23}
  {\path{doi:10.1007/978-3-642-03359-9_23}}.

\bibitem{Hotz65}
Günter Hotz.
\newblock Eine algebraisierung des syntheseproblems von schaltkreisen.
\newblock {\em EIK, Bd. 1, (185-205), Bd, 2, (209-231)}, 1965.

\bibitem{JOYAL91}
André Joyal and Ross Street.
\newblock The geometry of tensor calculus, i.
\newblock {\em Advances in Mathematics}, 88(1):55--112, 1991.
\newblock \href {https://doi.org/10.1016/0001-8708(91)90003-P}
  {\path{doi:10.1016/0001-8708(91)90003-P}}.

\bibitem{Kelly72}
G.~M. Kelly.
\newblock Many-variable functorial calculus. i.
\newblock In G.~M. Kelly, M.~Laplaza, G.~Lewis, and Saunders Mac~Lane, editors,
  {\em Coherence in Categories}, pages 66--105, Berlin, Heidelberg, 1972.
  Springer Berlin Heidelberg.
\newblock \href {https://doi.org/BFb0059556} {\path{doi:BFb0059556}}.

\bibitem{Kelly64}
Max Kelly.
\newblock On maclane’s conditions for coherence of natural associativities,
  commutativities, etc.
\newblock {\em Journal of Algebra}, 1:397--402, 1964.

\bibitem{Kinoshita97:ecategories}
Yoshiki Kinoshita.
\newblock A bicategorical analysis of e-categories, 02 1997.
\newblock URL:
  \url{https://www.researchgate.net/publication/2710125_A_bicategorical_analysis_of_E-categories}.

\bibitem{Quantomatic25}
Aleks Kissinger and Vladimir Zamdzhiev.
\newblock Quantomatic: A proof assistant for diagrammatic reasoning.
\newblock In Amy~P. Felty and Aart Middeldorp, editors, {\em Automated
  Deduction - CADE-25}, pages 326--336, Cham, 2015. Springer International
  Publishing.

\bibitem{soft:chyp}
Alekx Kissinger.
\newblock Chyp, 2023.
\newblock URL: \url{https://github.com/akissinger/chyp}.

\bibitem{lafontDiagramEditorMechanise2024}
Ambroise Lafont.
\newblock A diagram editor to mechanise categorical proofs.
\newblock In {\em 35es {{Journ{\'e}es Francophones}} Des {{Langages
  Applicatifs}} ({{JFLA}} 2024)}, Saint-Jacut-de-la-Mer, France, January 2024.
\newblock \url{https://amblafont.github.io/graph-editor}.
\newblock URL: \url{https://hal.science/hal-04407118}.

\bibitem{MacLane63}
Saunders~Mac Lane.
\newblock Natural associativity and commutativity.
\newblock {\em Rice University Studies}, 49:28--46, 1963.

\bibitem{lehmann2026vyzx}
Adrian Lehmann, Ben Caldwell, Bhakti Shah, William Spencer, and Robert Rand.
\newblock {VyZX}: Formal verification of a graphical quantum language.
\newblock {\em ACM Transactions on Programming Languages and Systems}, 2026.
\newblock To appear.
\newblock \href {https://arxiv.org/abs/2311.11571} {\path{arXiv:2311.11571}},
  \href {https://doi.org/10.48550/arXiv.2311.11571}
  {\path{doi:10.48550/arXiv.2311.11571}}.

\bibitem{MacLane:CWM}
Saunders Mac~Lane.
\newblock {\em Categories for the Working Mathematician}.
\newblock Graduate texts in mathematics 5. Springer, 2nd ed., reprint edition,
  1971 - 1998.

\bibitem{Penrose71}
Roger Penrose.
\newblock Applications of negative dimensional tensors.
\newblock {\em Combinatorial Mathematics and its Applications}, 1971.

\bibitem{Penrose_Rindler_1984}
Roger Penrose and Wolfgang Rindler.
\newblock {\em Spinors and Space-Time}.
\newblock Cambridge Monographs on Mathematical Physics. Cambridge University
  Press, 1984.
\newblock \href {https://doi.org/10.1017/CBO9780511564048}
  {\path{doi:10.1017/CBO9780511564048}}.

\bibitem{Piedeleu_Zanasi_2025}
Robin Piedeleu and Fabio Zanasi.
\newblock {\em An Introduction to String Diagrams for Computer Scientists}.
\newblock Elements in Applied Category Theory. Cambridge University Press,
  2025.
\newblock \href {https://doi.org/10.1017/9781009625715}
  {\path{doi:10.1017/9781009625715}}.

\bibitem{EilenBergKelly66}
G.~Max~Kelly Samuel~Eilenberg.
\newblock Closed categories.
\newblock In {\em Proceedings of the Conference on Categorical Algebra}, pages
  421--562. Springer, 1966.
\newblock Section II.1.
\newblock \href {https://doi.org/10.1007/978-3-642-99902-4}
  {\path{doi:10.1007/978-3-642-99902-4}}.

\bibitem{Shah_2025}
Bhakti Shah, Willam Spencer, Laura Zielinski, Ben Caldwell, Adrian Lehmann, and
  Robert Rand.
\newblock Vicar: Visualizing categories with automated rewriting in coq.
\newblock {\em Electronic Proceedings in Theoretical Computer Science},
  429:234–248, September 2025.
\newblock \href {https://doi.org/10.4204/eptcs.429.13}
  {\path{doi:10.4204/eptcs.429.13}}.

\bibitem{Street96}
Ross Street.
\newblock Handbook of algebra.
\newblock In M.~Hazewinkel, editor, {\em Categorical structures}. Elsevier,
  1996.

\bibitem{DubricDybjierScott98:pcategories}
Djordje \v{C}ubri\'{c}, Peter Dybjer, and Philip Scott.
\newblock Normalization and the yoneda embedding.
\newblock {\em Mathematical Structures in Computer Science}, 8(2):153–192,
  April 1998.
\newblock \href {https://doi.org/10.1017/S0960129597002508}
  {\path{doi:10.1017/S0960129597002508}}.

\end{thebibliography}

\end{document}